\documentstyle{mn}

\newif\ifAMStwofonts

\input epsf

\def\lesssim{\mathrel{\hbox{\rlap{\hbox{\lower4pt\hbox{$\sim$}}}\hbox{$<$}}}}

\def\gtrsim{\mathrel{\hbox{\rlap{\hbox{\lower4pt\hbox{$\sim$}}}\hbox{$>$}}}}

\def\msun{M$_{\odot}$}

\def\teff{T$_{\rm eff}$~}

\def\ll_lsun{$\log({L/\rm L_{\odot}})$~}

\def\masa_msun{$M/ \rm M_{\odot}$~}

\def\m_mstar{$M/M_{*}$~}

\title{Diffusion  and  the occurrence  of  hydrogen  shell flashes  in
helium white dwarf stars}

\author[L.  G. Althaus,  A.  M.  Serenelli and  O.  G. Benvenuto]  {L.
G.   Althaus\thanks{Member    of   the   Carrera    del   Investigador
Cient\'{\i}fico y  Tecnol\'ogico, Consejo Nacional  de Investigaciones
Cient\'{\i}ficas   y    T\'ecnicas   (CONICET),   Argentina.    Email:
althaus@fcaglp.fcaglp.unlp.edu.ar}  A.  M. Serenelli\thanks{Fellow  of
the Consejo Nacional  de Investigaciones Cient\'{\i}ficas y T\'ecnicas
(CONICET), Argentina.   Email: serenell@fcaglp.fcaglp.unlp.edu.ar} and
O.  G.   Benvenuto\thanks{Member  of  the   Carrera  del  Investigador
Cient\'{\i}fico, Comisi\'on de  Investigaciones Cient\'{\i}ficas de la
Provincia      de       Buenos      Aires,      Argentina.      Email:
obenvenuto@fcaglp.fcaglp.unlp.edu.ar}   \\    Facultad   de   Ciencias
Astron\'omicas  y Geof\'{\i}sicas, Universidad  Nacional de  La Plata,
Paseo del Bosque S/N, (1900) La Plata, Argentina}

\date{November 20}

\pagerange{\pageref{firstpage}--\pageref{lastpage}}

\pubyear{2000}

\begin{document}

\maketitle

\label{firstpage}

\begin{abstract}  In  this  paper  we  investigate the effects of
element diffusion  on the structure  and evolution of  low-mass helium
white  dwarfs.   Attention is  focused  mainly  on  the occurrence  of
hydrogen shell  flashes induced by diffusion  processes during cooling
phases. Physically sound initial  models with stellar masses of 0.406,
0.360,  0.327,  0.292, 0.242,  0.196,  0.169  and  0.161 \msun  \  are
constructed by applying mass loss rates at different stages of the red
giant branch  evolution of a  solar model up  to the moment  the model
begins  to  evolve   to  the  blue  part  of   the  HR  diagram.   The
multicomponent flow  equations describing gravitational  settling, and
chemical  and   thermal  diffusion   are  solved  and   the  diffusion
calculations are  coupled to an  evolutionary code.  In  addition, the
same sequences are computed but neglecting diffusion.  Results without
diffusion are similar to those of Driebe et al. (1998).

We  find that  element diffusion  strongly affects  the  structure and
cooling  history of  helium  white dwarfs.   In particular,  diffusion
induces the occurrence of hydrogen shell flashes in models with masses
ranging from 0.18  to 0.41 \msun, which is in  sharp contrast from the
situation when diffusion is  neglected. In connection with the further
evolution,  these  diffusion-induced  flashes  lead  to  much  thinner
hydrogen  envelopes, preventing  stable nuclear  burning from  being a
sizeable energy  source at advanced stages of  evolution. This implies
much  shorter  cooling  ages  than  in  the  case  when  diffusion  is
neglected.

These  new  evolutionary  models  are  discussed in  light  of  recent
observational data of some millisecond pulsar systems with white dwarf
companions.  We find that age discrepancies between the predictions of
standard evolutionary  models and such  observations appear to  be the
result  of ignoring  element  diffusion in  such evolutionary  models.
Indeed, such discrepancies vanish when account is made of diffusion.

\end{abstract}


\section{Introduction} \label{sec:intro}

Over the last  few years, low-mass helium white  dwarf (WD) stars have
been  detected  in numerous  binary  configurations.  The  theoretical
prediction that low  mass WDs with helium core would  be the result of
mass transfer episodes in close binary systems (Paczy\'nski 1976; Iben
\&  Webbink 1989;  Iben  \& Livio  1993)  was first  place  on a  firm
observational  ground  by  Marsh  1995  and  Marsh,  Dhillon  \&  Duck
1995. From then on, these objects have been observed in various binary
systems containing usually  either another WD or a  neutron star (see,
e.g., Lundgren et  al. 1996; Moran, Marsh \&  Bragaglia 1997; Orosz et
al. 1999; Maxted  et al.  2000). In addition,  several helium WDs have
been found  in open and globular  clusters (Landsman et  al.  1997 and
Edmonds et  al.  1999).   Evolutionary models for  these WDs  with the
emphasis on their mass-radius relations have been presented by Althaus
\& Benvenuto  (1997), Benvenuto \&  Althaus (1998), Hansen  \& Phinney
(1998a) and Driebe et al.(1998).

An important  configuration in  which helium WDs  can be  expected are
those binary systems involving the presence of millisecond pulsars. In
this  regard, numerous observations  reveal that  the majority  of WDs
having  millisecond pulsar  companions  are low  mass  WDs (Hansen  \&
Phinney 1998b and references cited  therein; van Kerkwijk et al. 2000;
and  Phinney  \& Kulkarni  1994  for  a  review). In  particular,  the
presence  of a WD  in these  binary systems  offers an  opportunity to
check assumptions made about the ages of millisecond pulsar.  In fact,
cooling ages for  WDs provide an estimation of the  age of the system,
independent of  spin-down age of the  pulsar.  This is  so because the
pulsar  spin down  begins at  nearly the  same time  as  the companion
adopts a WD  configuration, after the end of  mass loss episodes.  The
PSR  J1012+5307  system is  the  best  studied  system of  this  type.
Indeed,   this  system   has  captured   the  attention   of  numerous
investigators and atmospheric parameters  of the low-mass WD companion
are relatively  well known (van  Kerkwijk, Bergeron \&  Kulkarni 1996;
Callanan, Garnavich \& Koester 1998). In particular, the spin-down age
of the pulsar is approximately 7 Gyr (Lorimer et al. 1995). This value
for the pulsar age can be compared with theoretical predictions of the
cooling of its helium WD companion.

In  this connection, Sarna,  Antipova \&  Ergma (1999)  have presented
very  detailed calculations  of the  binary evolution  leading  to the
formation of low-mass,  helium WDs with stellar masses  less than 0.25
\msun. They  found that,  after detachment of  the Roche  lobe, helium
cores are surrounded by a massive  hydrogen layer of 0.01 - 0.06 \msun
\ with  a surface hydrogen  abundance by mass  of $X_{\rm H}=$  0.35 -
0.50.  Massive  hydrogen  envelopes  for  helium WDs  have  also  been
derived  by  Driebe  et  al.(1998),  who  have  simulated  the  binary
evolution by forcing  a 1 \msun \  model at the red giant  branch to a
large  mass  loss rate.  Driebe  et  al.  (1998) have  followed  their
calculations  down  to  very  low  stellar  luminosities  and  derived
mass-radius relations  in order to analyse  recent observational data.
These studies  (see also  Webbink 1975 and  Alberts et al.  1996) show
that  thermal  flashes  do  not   occur  for  WDs  less  massive  than
approximately  0.2  \msun  \  and  that  the  predicted  hydrogen-rich
envelope is  massive enough so that stationary  hydrogen shell burning
remains  dominant for helium  WDs even  down to  effective temperature
(\teff)  values  below  10000K,  thus substantially  prolonging  their
cooling times.   In particular, these evolutionary models  give an age
for  the  helium WD  in  PSR J1012+5307  in  good  agreement with  the
spin-down age of the pulsar.

However, recent observational  studies seem to cast some  doubt on the
thickness  of the  hydrogen  envelopes predicted  by the  evolutionary
calculations  mentioned  in  the  foregoing  paragraph.   Indeed,  van
Kerkwijk  et  al.   (2000)  have  detected the  WD  companion  to  the
millisecond pulsar  PSR B1855+09 and  determined its \teff to  be 4800
$\pm$ 800K.   Since the mass of  the WD is accurately  known thanks to
the      Shapiro       delay      of      the       pulsar      signal
(0.258$^{+0.028}_{-0.016}$\msun; see Kaspi, Taylor \& Ryba 1994), this
low \teff value corresponds to a WD cooling age of 10 Gyr according to
the  Driebe et  al. (1998)  evolutionary  models.  This  is in  strong
discrepancy  with the characteristic  age of  the pulsar,  $\tau_c=$ 5
Gyr.   Interestingly  enough,  the  pulsars  PSR  J0034-0534  and  PSR
J1713+0747  have very cool  WD companions  (Hansen \&  Phinney 1998b),
thus  implying cooling  ages far  grater than  10 Gyr  from  Driebe et
al.  models.  As  discussed  by  van Kerkwijk  et  al.  (2000),  these
evolutionary  models  may  overestimate  the cooling  ages.   Finally,
Ergma, Sarna  \& Antipova (2000)  have recently discussed  the cooling
history of PSR  J0751+1807 and found that helium  WD models with thick
hydrogen  envelopes implies cooling  ages too  high for  the estimated
\teff of the WD component.

Needless to say, the cooling age of  the WD depends on the mass of the
hydrogen envelope  before entering the  cooling track.  In  fact, less
massive hydrogen envelopes prevent hydrogen burning from being a major
source of energy  during the further evolution of  the WD, thus giving
rise to low  ages over late cooling stages.   There exist several ways
by  which the  mass  of the  hydrogen  envelope can  be reduced.   For
instance, it  is possible that after  a flash episode  the WD envelope
reaches again  giant dimensions,  and another mass  loss phase  can be
initiated, causing a substantial reduction of the mass of the hydrogen
layers. This possibility has been  analysed by Iben \& Tutukov (1986),
who  have presented a  detailed treatment  of the  evolution of  a 0.3
\msun \ helium WD remnant in  a binary system. These authors find that
their  model experiences two  hydrogen shell  flashes which  force the
model to  again fill  its Roche lobe.   As a  result, the mass  of the
hydrogen envelope  is considerably reduced ($M_H=  1.4 \times 10^{-4}$
\msun), which leads to small cooling ages at late stages of evolution.
In this sense  the existence of hydrogen shell  flashes should play an
important role in determining the thickness of the hydrogen layer atop
a helium  WD, and thus in  establishing the actual  time-scale for the
further evolution  of the star. Another possibility  has recently been
explored by  Ergma et al.  (2000),  who propose that a  very low mass,
helium  pre-WD,   after  detachment  of  its  Roche   lobe,  looses  a
considerable  fraction   of  its  hydrogen  envelope   due  to  pulsar
irradiation.
  
Here, we  undertake the current  investigation in order to  assess the
role  played by  diffusion in  the occurrence  of hydrogen  flashes in
helium  WDs and  more importantly  to investigate  whether or  not the
hydrogen  envelope  mass  can  be  considerably  reduced  by  enhanced
hydrogen  burning  during  flash   episodes.  As  a  matter  of  fact,
gravitational settling  and chemical diffusion are known  to alter the
distributions  of element  abundances below  the stellar  surface.  In
particular,  chemical diffusion  is  expected  to lead  to  a tail  of
hydrogen  penetrating   downwards  through  hotter   layers  (Iben  \&
MacDonald 1985  and Althaus \& Benvenuto  2000). In the case  of a 0.3
\msun  \  helium WD,  this  feature has  been  suspected  to induce  a
hydrogen  shell flash  (Iben \&  Tutukov 1986),  forcing the  model to
reach giant dimensions again.

Diffusion  processes in WDs  have captured  the attention  of numerous
investigators since  the early  work by Schatzman  (1958) on  the role
played  by diffusion  in  the evolution  of  the superficial  chemical
composition of  WDs.  From  then on, various  studies have  shown that
diffusion processes  cause elements heavier than  the main atmospheric
constituents  to sink  below  the photosphere  over  time scales  much
smaller  than   the  evolutionary  time  scales,   thus  providing  an
explanation for the  purity of almost all WD  atmospheres (Fontaine \&
Michaud  1979;  Alcock \&  Illarionov  1980;  Muchmore  1984; Iben  \&
MacDonald 1985;  Paquette et al.  1986b; Dupuis et al.   1992, amongst
others).   In  addition, sophisticated  models  invoking an  interplay
amongst  various mechanisms such  as diffusion,  convection, accretion
and  wind mass loss  have been  developed to  explain the  presence of
heavy elements in  very small proportions detected in  the spectrum of
some WDs (see,  e.g., Fontaine \& Michaud 1979;  Vauclair, Vauclair \&
Greenstein 1979; Pelletier et al. 1986; Paquette et al. 1986b; Iben \&
MacDonald 1985;  Dupuis et  al. 1992; Unglaub  \& Bues 1998).   In the
context of helium WD evolution, Iben \& Tutukov (1986) have found that
diffusion processes  carry some hydrogen inwards  through hotter layer
to such an  extent that a second hydrogen shell flash  is induced in a
0.3 \msun  \ model.  Very  recently, Althaus \& Benvenuto  (2000) have
explored  the effect  of  diffusion in  the  mass-radius relation  for
helium WDs.

In  this work,  we explore  the  evolution of  helium WDs  in a  self-
consistent way with the evolution of element abundances resulting from
diffusion  processes.  The  diffusion  calculations are  based on  the
multicomponent treatment  of the gas  developed by Burgers  (1969) and
gravitational settling, thermal  and chemical diffusion are considered
in  our  calculations.   Reliable  initial models  are  obtained  from
explicit modeling of their pre-WD evolution.  To this end, we simulate
the mass exchange  phases expected in a real  situation by forcing the
evolution  of a  1 \msun  \ model  to a  sufficiently large  mass loss
rates.  We shall see that diffusion is a fundamental ingredient in the
evolution  of  helium WDs  that  has to  be  taken  into account  when
assessing  general  characteristics  of  binary systems  containing  a
helium WD.  In particular, we shall see that several aspects of helium
WD evolution are strongly altered, as compared with the situation when
diffusion is  neglected (Driebe et al. 1998).   Amongst these aspects,
we  mention  the  fact  that  nuclear  burning  remains  insignificant
throughout  the  entire  cooling  phase  following the  end  of  flash
episodes, thus resulting in  considerably shorter evolutionary ages of
helium WDs.  Needless to say, this feature  has important implications
when attempt  is made in  comparing theoretical predictions on  the WD
evolution  with expectations from  millisecond pulsars.  Details about
our  evolutionary code,  diffusion  treatment and  initial models  are
given in  Section 2.  Results are presented  in Section 3  and finally
Section 4 is devoted to making some concluding remarks.

\section{Computational details}

\subsection{Evolutionary code and diffusion equations}

The  evolutionary code  we used  in the  present calculation  has been
described in our previous works  on WD evolution (Althaus \& Benvenuto
1997,  1998)   and  we  refer   the  reader  to  those   articles  for
details. Briefly,  our code  is based on  a very detailed  and updated
physical  description such  as OPAL  radiative opacities  (Iglesias \&
Rogers 1996) and molecular opacities (Alexander \& Ferguson 1994); the
equation of state for the low  density regime is an updated version of
that of Magni \& Mazzitelli  (1979), while for the high density regime
we  consider  ionic  contributions,  Coulomb  interactions,  partially
degenerate   electrons,  and   electron   exchange  and   Thomas-Fermi
contributions at finite temperature. High-density conductive opacities
and the  various mechanisms of  neutrinos emission are taken  from the
works of  Itoh and  collaborators (see Althaus  \& Benvenuto  1997 for
details).   Hydrogen burning is  taken into  account by  considering a
complete network of thermonuclear  reaction rates corresponding to the
proton-proton chain and the  CNO bi-cycle.  Nuclear reaction rates are
taken from Caughlan \& Fowler (1988). Electron screening is treated as
in Wallace, Woosley \& Weaver (1982).

Gravitational settling,  and chemical and thermal  diffusion have been
fully taken  into account.  To this  end, we follow  the treatment for
multicomponent gases  presented by  Burgers (1969), thus  avoiding the
trace element approximation usually invoked  in most WD studies. It is
worth noting that after mass  loss episodes helium WDs are expected to
have envelopes made  up of a mixture of hydrogen  and helium, thus the
trace element  approximation would clearly not be  appropriate for the
case we want to study  here.  Radiative levitation, which is important
for  determining  photospheric composition  of  hot  WDs (Fontaine  \&
Michaud 1979)  has been neglected. This assumption  is justified since
we are  interested in the  chemical evolution occurring quite  deep in
the star. In the context  of WD evolution, the treatment for diffusion
we use here has been employed by Muchmore (1984) and Iben \& MacDonald
(1985). Recently, it has been  applied by MacDonald, Hernanz \& Jos\'e
(1999) to  address  the  problem  of  carbon  dredge-up  in  WDs  with
helium-rich envelopes.

Under the  influence of  gravity, partial pressure,  thermal gradients
and induced electric fields  (we neglect stellar rotation and magnetic
fields) the  diffusion velocities  in a multicomponent  plasma satisfy
the  set of  diffusion equation  ($N-1$ independent  linear equations,
Burgers 1969)

\begin{eqnarray}
{{dp_i} \over {dr}}-{{\rho  _i} \over \rho}{{dp} \over {dr}}-n_iZ_ieE=
\sum\limits_{j\ne   i}^{N}  {K_{ij}}\left({w_j-w_i}  \right)   \cr  +\
\sum\limits_{j\ne   i}^{N}  {K_{ij}\  z_{ij}}   {{m_j\  r_i\   -  m_i\
r_j}\over{m_i\ + m_j}},
\end{eqnarray}

\noindent and heat flow equation ($N$ equations)

\[{{5} \over {2}}n_i k_{B} \nabla T= - {{5} \over {2}}
\sum\limits_{j\ne   i}^{N}   {K_{ij}\   z_{ij}}   {{m_j}\over{m_i\   +
m_j}}\left({w_j-w_i} \right)  - {{2} \over  {5}}{K_{ii}}\ z_{ii}^{,,}\
r_i \]
\begin{eqnarray}
 -\sum\limits_{j\ne  i}^{N} {{K_{ij}}\over{(m_i\  +  m_j)^2}}\left( {3
m_i^2  + m_j^2  z_{ij}^{,}  + 0.8m_im_jz_{ij}^{,,}}  \right)\ r_i  \cr
{+\sum\limits_{j\ne i}^{N} {{K_{ij}m_im_j}\over{(m_i\ + m_j)^2}}\left(
{3 + z_{ij}^{,} - 0.8z_{ij}^{,,}} \right)\ r_j}.
\end{eqnarray}

Here, $p_i$, $\rho_i$, $n_i$, $Z_i$ and $m_i$ means, respectively, the
partial pressure,  mass density, number density, mean  charge and mass
for  species  $i$  ($N$  means   the  number  of  ionic  species  plus
electron). $T$, $k_{B}$ and  $\nabla T$ are the temperature, Boltzmann
constant and temperature gradient, respectively. The unknown variables
are  the diffusion  velocities with  respect  to the  center of  mass,
$w_i$, and the residual heat flows $r_i$ (for ions and electrons).  In
addition the electric field $E$  has to be determined.  The resistance
coefficients ($K_{ij}, z_{ij}, z_{ij}^{,}$ and $z_{ij}^{,,}$) are from
Paquette  at  al  (1986a)  and  averages  ionic  charges  are  treated
following  an  approximate  pressure  ionization  model  as  given  by
Paquette et al. (1986b), which is sufficient for our purposes.

To complete  the set of  equations, we use  the conditions for  no net
mass flow with respect to the center of mass

\begin{eqnarray}
\sum\limits_{i} {A_i} n_i w_i=0,
\end{eqnarray}

\noindent and no electrical current

\begin{eqnarray}
\sum\limits_{i} {Z_i} n_i w_i=0.
\end{eqnarray}

\noindent  In terms  of  the gradient  in  the number  density we  can
transform Eq. (1) to

\begin{eqnarray}
{{1}\over{n_i}}\left[\sum\limits_{j\ne  i}^{N} {K_{ij}}\left({w_i-w_j}
\right)+\ \sum\limits_{j\ne i}^{N} {K_{ij}\ z_{ij}} {{m_i\ r_j\ - m_j\
r_i}\over{m_i\  +  m_j}}\right] \cr  -\  Z_ieE=  \alpha_i  - k_{B}  T\
{{d\ln{n_i}}\over{dr}},
\end{eqnarray}

\noindent where

\begin{eqnarray}
\alpha_i = - A_i m_H g - k_{B} T\ {{d\ln{T}}\over{dr}},
\end{eqnarray}

\noindent where $A_i$, $m_H$, $g$  and $T$ are the atomic mass number,
hydrogen  atom mass,  gravity  and temperature,  respectively. Let  us
write the  unknowns $w_i$, $r_i$ and  $E$ in terms of  the gradient of
ion densities in the form (similarly for $r_i$ and $E$)

\begin{eqnarray}
w_i    =   w_{i}^{gt}    -    \sum\limits_{ions   (j)}    \sigma_{ij}\
{{d\ln{n_j}}\over{dr}},
\end{eqnarray}

\noindent where $w_{i}^{gt}$ stands  for the velocity component due to
gravitational  settling  and  thermal  diffusion.   The  summation  in
Eq. (7) is to  be effected over the ions only.  With  Eqs. (2) and (5)
together  with  (3)  and  (4)   we  can  easily  find  the  components
$w_{i}^{gt}$    and   $\sigma_{ij}$    by   matrix    inversions   (LU
decomposition).

Now,  we  are  in  a   position  to  find  the  evolution  of  element
distribution   throughout   the  star   by   solving  the   continuity
equation. Details are  given in Althaus \& Benvenuto  (2000) (see also
Iben \&  MacDonald 1985).  In  particular, we follow the  evolution of
the  isotopes  $^{1}$H,  $^{3}$He,  $^{4}$He, $^{12}$C,  $^{14}$N  and
$^{16}$O.  In  order to calculate  the dependence of the  structure of
our WD models on the  varying abundances self-consistently, the set of
equations describing  diffusion has  been coupled to  our evolutionary
code.   We  want  to  mention  that  after  computing  the  change  of
abundances  by effect  of  diffusion, they  are  evolved according  to
nuclear reactions and convective  mixing.  It is worth mentioning that
radiative opacities  are calculated for  metallicities consistent with
diffusion predictions. In particular,  the metallicity is taken as two
times the abundances of CNO elements.

\begin{table}
\centering
\begin{minipage}{60mm}
\caption{Stellar  mass,  envelope   mass  and  hydrogen  surface  mass
fraction $X_H$ at the point of maximum effective temperature.}

\begin{tabular*}{60mm}{@{\extracolsep{\fill}}ccc@{}}
\hline  $M/{\rm M_{\odot}}$  &  $M_{env}/{\rm M_{\odot}}$  & $X_H$  \\
\hline 0.406 & 7.3  $ \times 10^{-4}$ & 0.701 \\ 0.360  & 1.1 $ \times
10^{-3}$ & 0.701 \\  0.327 & 1.4 $ \times 10^{-3}$ &  0.701 \\ 0.292 &
2.0 $ \times 10^{-3}$ & 0.701 \\ 0.242 & 3.7 $ \times 10^{-3}$ & 0.694
\\ 0.196  & 6.7  $ \times  10^{-3}$ & 0.504  \\ 0.169  & 1.0  $ \times
10^{-2}$ & 0.423 \\ \hline
\end{tabular*}
{\small  The  quoted  values  correspond  to the  sequences  in  which
diffusion is neglected}
\medskip
\end{minipage}
\end{table}

\subsection{Initial models}

\begin{figure}
\epsfxsize=240pt \epsfbox[19 190 578 800]{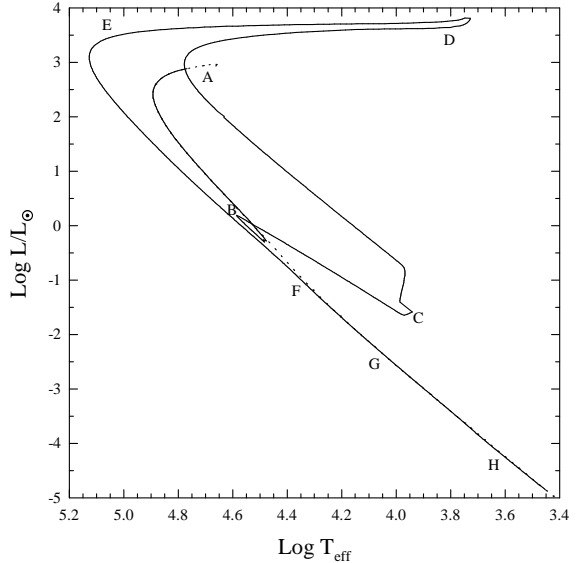}
\vskip -1.4cm
\caption{ Hertzsprung - Russell (HR)  diagram for the evolution of the
0.406 $M_{\odot}$ helium WD model.  Solid line corresponds to the case
when diffusion is included and  dotted line depicts the situation when
diffusion  is neglected.  Note that  diffusion processes  lead  to the
occurrence of a hydrogen shell flash, forcing the model to evolve back
to the red giant region (point  D). This is in sharp contrast with the
situation encountered  under the assumption of no  diffusion, in which
case flash episodes do not occur at all. Characteristics of  the
models with diffusion labeled by letters are given in Table 2.}

\end{figure}

\begin{figure}
\epsfxsize=240pt \epsfbox[19 190 578 800]{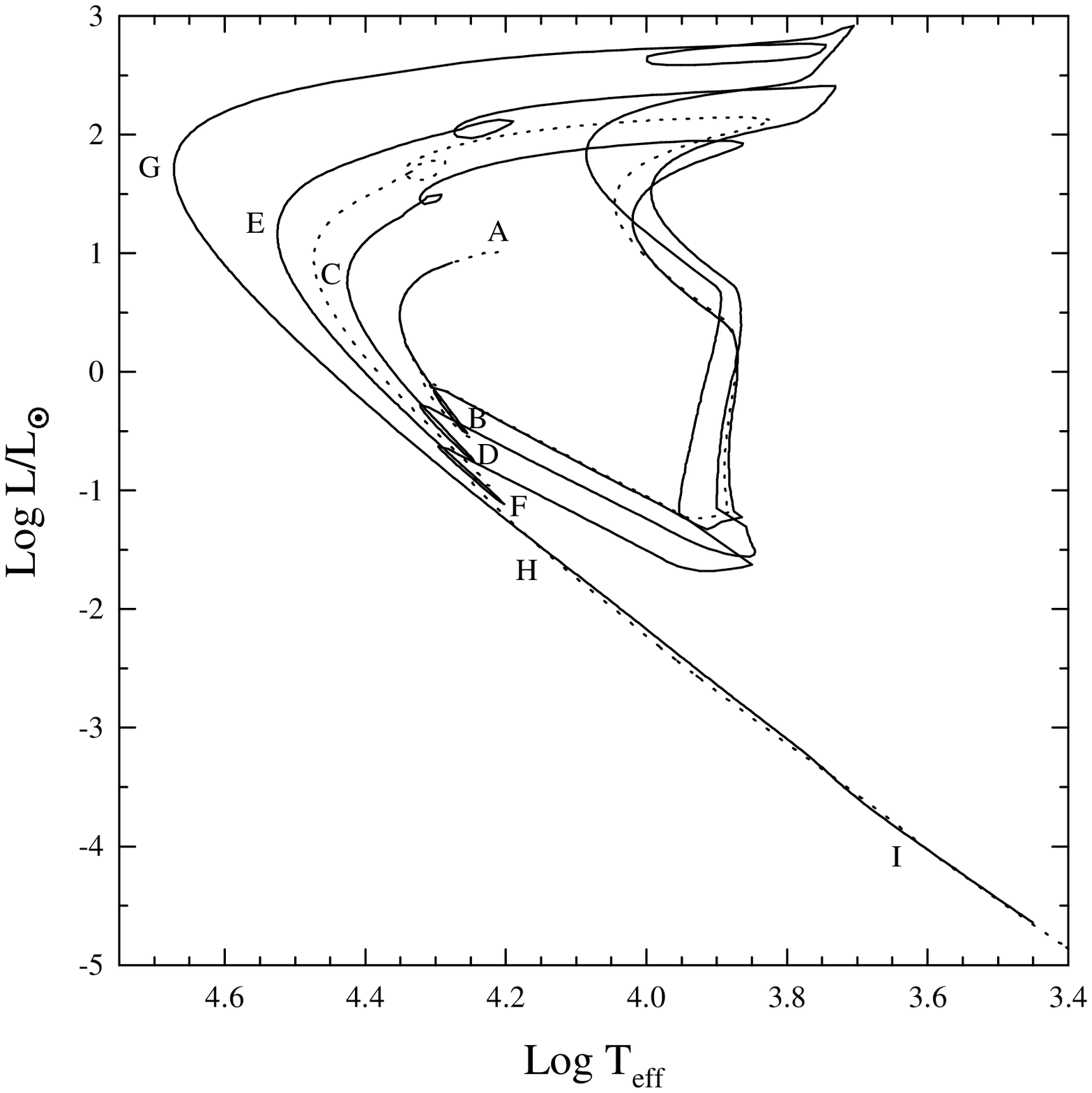}
\vskip -1.4cm
\caption{Same  as   Fig.  1  but  for  0.242   $M_{\odot}$  helium  WD
models.  Here diffusion gives  rise to  three hydrogen  shell flashes,
whilst   one  flash   is   obtained  when   diffusion  is   neglected.
Characteristics of  the models with  diffusion labeled by  letters are
given in Table 2. }
\end{figure}

\begin{figure}
\epsfxsize=240pt \epsfbox[19 190 578 800]{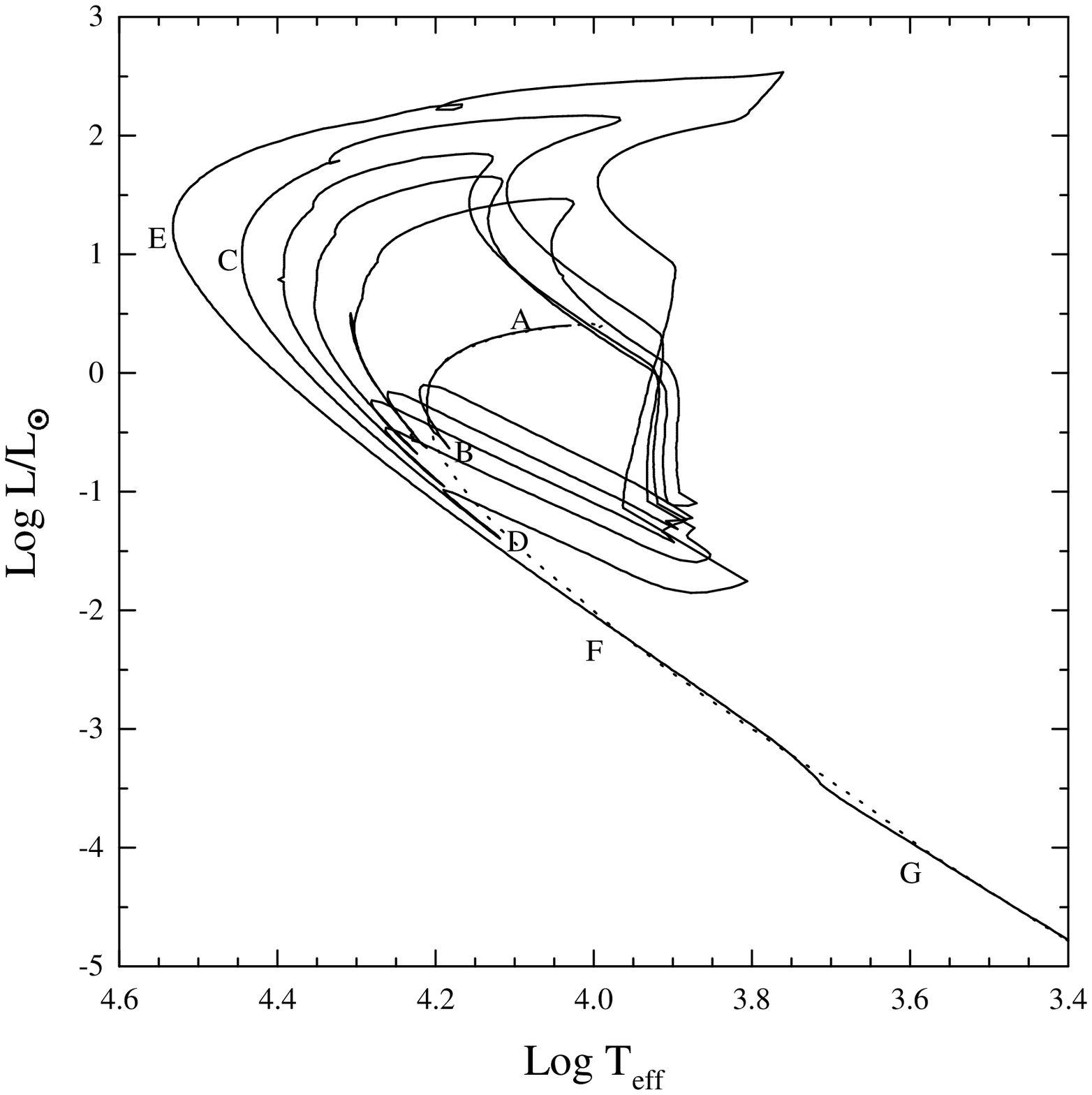}
\vskip -1.35cm
\caption{Same  as   Fig.  1  but  for  0.196   $M_{\odot}$  helium  WD
models.  Here diffusion  gives rise  to five  hydrogen  shell flashes,
whilst   no   flash  is   obtained   when   diffusion  is   neglected.
Characteristics of  the models with  diffusion labeled by  letters are
given in Table 2.}
\end{figure}

\begin{figure}
\epsfxsize=240pt \epsfbox[19 150 578 800]{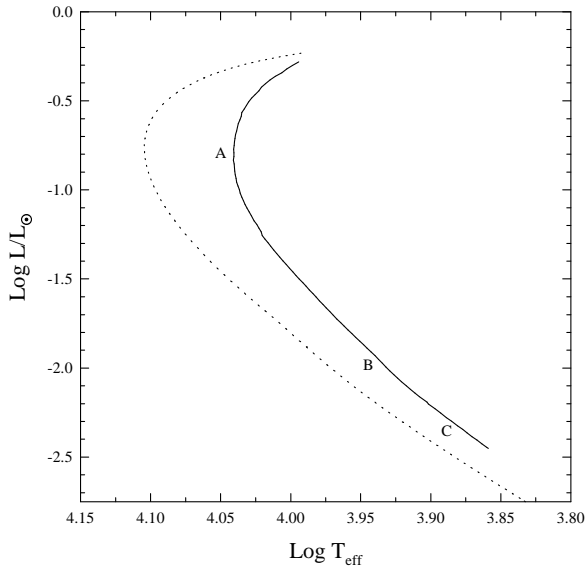}
\vskip -2cm
\caption{Same as  Fig. 1 but  for 0.169 $M_{\odot}$ helium  WD models.
Unlike more massive  models, here no hydrogen shell  flash takes place
even  in the  presence of  diffusion.  This  results in  quite similar
evolutionary ages  for both sets  of sequences.  However note  that at
low  $T_{\rm eff}$ values,  models become  markedly less  compact when
diffusion in considered. Characteristics  of the models with diffusion
labeled by letters are given in Table 2.}
\end{figure}

\begin{figure}
\epsfxsize=240pt \epsfbox[19 210 578 800]{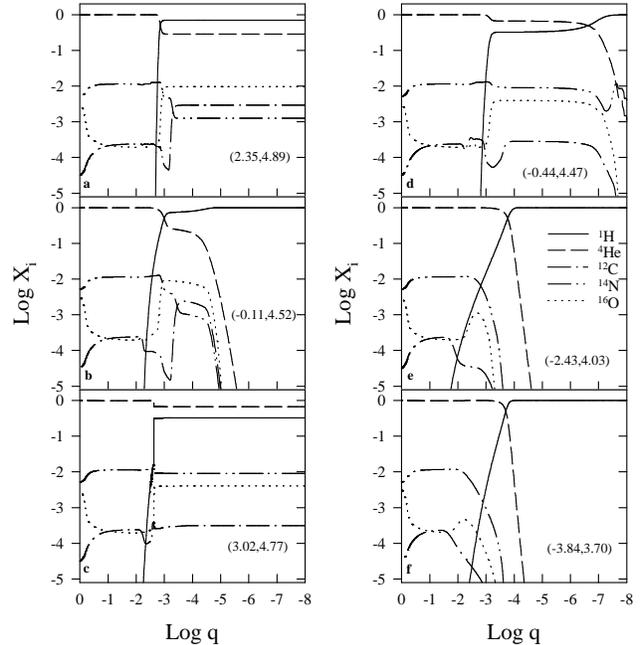}
\caption{Abundance by  mass of  $^1$H, $^4$He, $^{12}$C,  $^{14}$N and
$^{16}$O versus the outer  mass fraction $q$ ($q=1-M_r/M_*$) for 0.406
$M_{\odot}$   helium  WD  models   at  selected   evolutionary  stages
characterized  by values of  $\log{L/\rm L_{\odot}}$  and $\log{T_{\rm
eff}}$  (numbers given  between brackets).  Figure {\bf  a}  shows the
initial chemical  stratification before the model  reaches the cooling
branch for the first time. Figure {\bf b} corresponds to the situation
short before  the occurrence  of the hydrogen  shell flash  and figure
{\bf c}  depicts the stratification immediately  after mixing episodes
and  before the  model  returns  to the  red  giant region.   Finally,
figures {\bf  d}, {\bf e} and  {\bf f} depict  models corresponding to
the final  cooling branch.   Note that diffusion  substantially alters
the  chemical profiles,  rapidly leading  to pure  hydrogen envelopes.
Note also the  tail of hydrogen extending into  hotter and helium-rich
regions as  a result of chemical  diffusion (figures {\bf  b} and {\bf
e}).  At high $T_{\rm eff}$ values, this effect is responsible for the
occurrence of hydrogen shell flash episodes.}
\end{figure}

\begin{figure}
\epsfxsize=240pt \epsfbox[19 210 578 800]{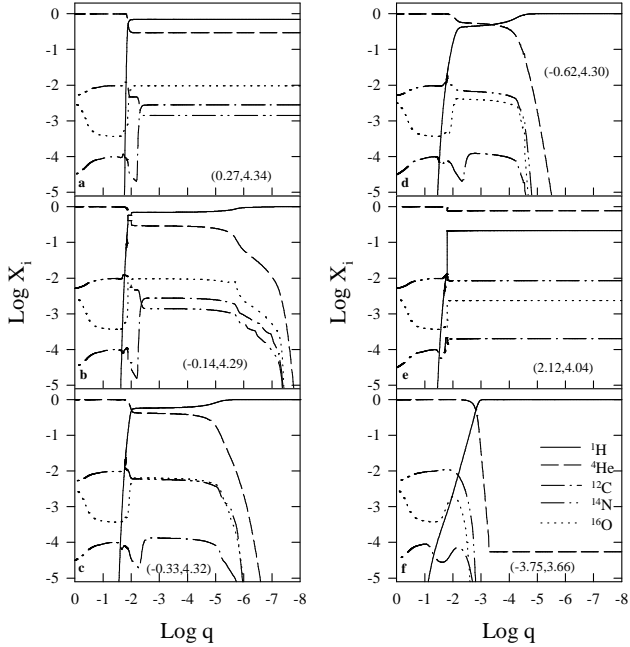}
\caption{Same as  Fig. 5 but  for 0.242 $M_{\odot}$ helium  WD models.
Here, figures {\bf b}, {\bf  c} and {\bf d} correspond to evolutionary
phases just before  the occurrence of each flash  episode. Figure {\bf
e} shows the chemical stratification immediately after mixing episodes
and before  the model  returns to  the red giant  region for  the last
time.  Finally, figure  {\bf f}  corresponds to  an advanced  stage of
evolution where some helium is present in the outer layers as a result
of convective mixing.  See text for details.}
\end{figure}

\begin{figure}
\epsfxsize=240pt \epsfbox[19 210 578 800]{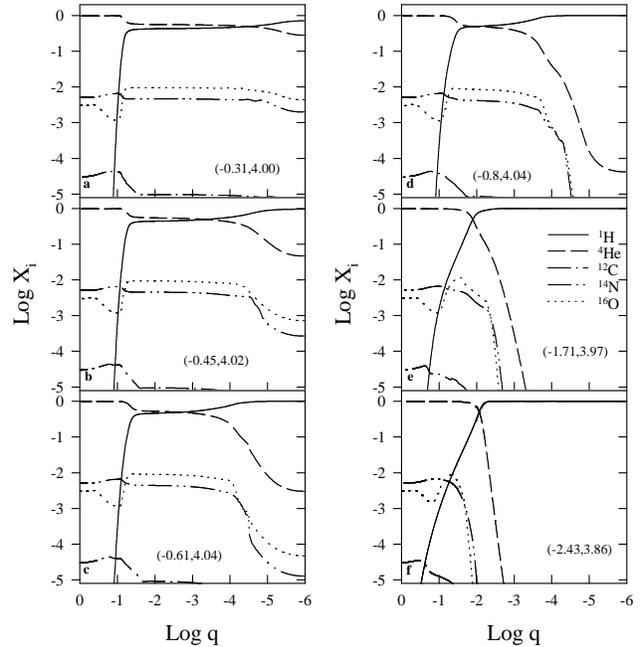}
\caption{Same as  Fig. 5 but  for 0.169 $M_{\odot}$ helium  WD models.
Figure {\bf  a} depicts the  chemical stratification before  the model
reaches  the cooling  branch, while  the other  figures  correspond to
models on the  cooling branch.  Note that diffusion  is very efficient
in modifying  the outer layer  composition even at high  $T_{\rm eff}$
values  due  to  the  large  evolutionary  times  characterizing  this
evolutionary phase.}
\end{figure}

\begin{figure}
\epsfxsize=240pt \epsfbox[19 210 578 800]{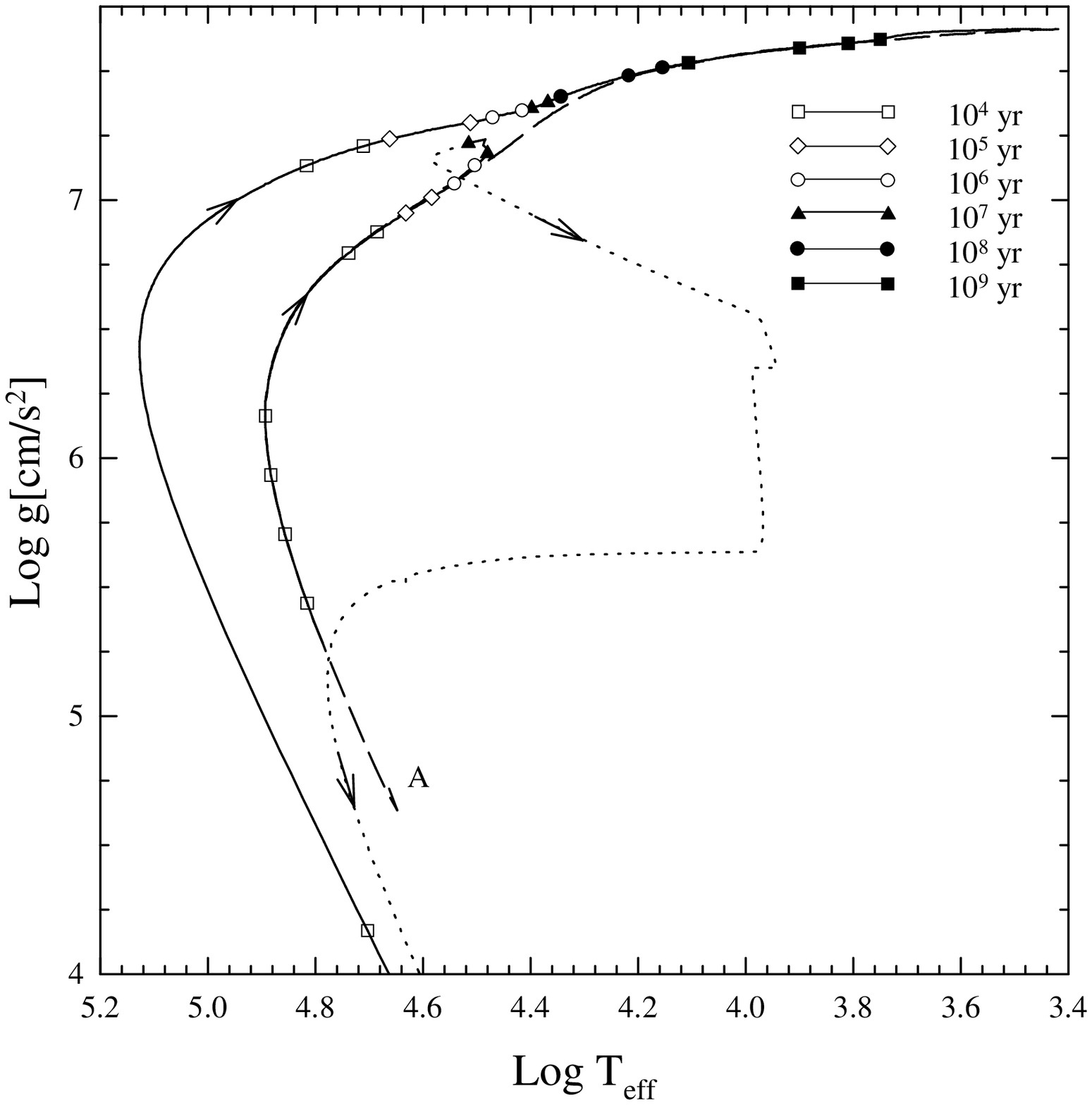}
\vskip -1cm
\caption{Surface   gravity  vs.    effective  temperature   for  0.406
$M_{\odot}$ helium WD models. Solid  line corresponds to the case when
diffusion is included and dashed  line to the situation when diffusion
is  neglected.  Both  set of  sequences start  at point  A.   The fast
evolution during the  flash phase is shown by  dotted lines and arrows
indicate the  course of evolution.  Also,  evolutionary time intervals
for models with  diffusion are indicated by means  of adjacent symbols
along  the curves.  The  effect of  the thermonuclear  flash resulting
from  element diffusion  upon surface  gravities is  apparent  at high
$T_{\rm  eff}$  values.   Except  for  such  stages,  where  evolution
proceeds very fast, surfaces  gravities are not appreciably changed by
the inclusion of diffusion processes.  See text for details.}
\end{figure}

In  this work  we  have not  modeled  in detail  the binary  evolution
leading to  the formation of  helium WDs (see  Sarna et al.   1999 and
references therein for details).  Rather, we have obtained our initial
models  by  simply  abstracting  mass  from  a  1  \msun  \  model  at
appropriate stages of its evolution  towards the red giant branch (see
also Iben \&  Tutukov 1986 and Driebe et al.  1998).   In this way, we
were able to generate initial  WD models with stellar masses of 0.406,
0.360, 0.327, 0.292, 0.242, 0.196  and 0.169 \msun. We want to mention
that the evolution  of the 1 \msun \ model has  been calculated on the
basis of a very detailed physical description, such as the equation of
state  of OPAL (Rogers,  Swenson \&  Iglesias 1996).   In Table  1, we
tabulate the  main characteristic of  our initial models. We  list, in
particular, for each  stellar mass, the envelope mass  and the surface
abundance  of hydrogen  at  maximum \teff.   Note  that our  resulting
envelopes and  hydrogen surface abundance  are in good  agreement with
those quoted by Driebe et  al (1998), particularly for massive models.
For less  massive models, however,  our envelopes are  somewhat larger
than  that  of  the   mentioned  authors.  Interestingly,  very  large
envelopes have  recently been  derived by Sarna  et al. (1999)  on the
basis of detailed binary evolution calculations.

We  want to  stress here  that the  approach we  follow to  obtain our
initial models  represents a simplification as compared  with the real
situation  in which  a common  envelope phase  is  expected.  However,
after the termination of mass  loss episodes, the further evolution of
the model does  not depend on the details of how  most of the envelope
was lost. In  particular, the mechanical and thermal  structure of the
models  are  consistent  with  the  predictions  of  binary  evolution
calculations (see Driebe et al. 1998)
 
\begin{table*}
\centering
\begin{minipage}{120mm}
\caption{Selected  stages for 0.406,  0.242, 0.196  and 0.169  \msun \
helium WD models considering diffusion}
\begin{tabular}{@{}cccccccc@{}}
\hline $M_*/{\rm  M_{\odot}}$ &  $Log(L{\rm /L}_\odot$) &  $Log T_{\rm
eff}$  &  $Age$ ($10^6$  yr)  & $Log  (g)$  &  $Log (L_{\rm  nuc}/{\rm
L_{\odot}})$ &  $X_H$ &  $Log (M_H/{\rm M_*})$  \\ \hline 0.406  (A) &
2.9465 & 4.7001 & 0.050 & 4.8559  & 2.9426 & 0.7007 & -2.764 \\ '' (B)
& 0.4136 &  4.6003 & 0.208 &  6.9897 &-0.6597 & 0.9690 &  -2.990 \\ ''
(C) &-1.6085 & 3.950 & 13.47880&  6.4105 & 6.2738 & 0.9999 & -3.112 \\
'' (D) & 3.6399 & 3.800 &  13.47882& 0.5621 & 3.4811 & 0.3217 & -3.118
\\ '' (E)  & 3.4490 & 5.0547 &  13.48295 & 5.7717 & 3.4427  & 0.3217 &
-3.454 \\ '' (F) &-0.7533 & 4.400 & 16.782 & 7.3555 & -2.4647 & 0.9997
& -3.599  \\ '' (G)  &-2.1305 &  4.100 & 351.53  & 7.5339 &  -3.7283 &
0.9999 & -3.628\\ '' (H) &-4.0503 & 3.650 & 8334.90 & 7.6527 & -7.5068
& 1.0000 &  -3.634\\ \hline \\ 0.242  (A) & 1.0199 & 4.2099  & 17.00 &
4.5962  & 1.0193  & 0.6936  & -1.892  \\ ''  (B) &-0.5299  &  4.2538 &
33.057& 6.3219 &-0.6130 & 0.9999 &  -2.035 \\ '' (C) & 0.7534 & 4.4258
& 40.209  & 5.7263 &  0.7457 &  0.9996 & -2.155  \\ '' (D)  &-0.7527 &
4.2454 & 45.827 & 6.5107 &-0.8618 & 1.0000 & -2.168 \\ '' (E) & 1.0765
&  4.5238 &  54.70 &  5.7951 &  1.0684  & 0.9820  & -2.347  \\ ''  (F)
&-1.1160 & 4.2025&  71.695 & 6.7027 & -1.2617 & 0.9997  & -2.363 \\ ''
(G) & 1.7033 &  4.6722& 95.974 & 5.7620 & 1.6940 &  0.4798 & -2.676 \\
''  (H) &-1.3322  &  4.1809& 100.59  &  6.8326 &  -1.9306  & 0.9997  &
-2.725\\ '' (I) &-3.8149 & 3.6506&  4222.8 & 7.1942 & -5.8365 & 0.9997
& -2.743\\ \hline  \\ 0.196 (A) &  0.3393 & 4.1000 & 39.53  & 4.7470 &
0.3367 & 0.5042 & -1.679 \\  '' (B) &-0.6347 & 4.1828 & 235.43& 6.0509
&-0.6521 &  1.0000 & -1.910  \\ ''  (C) & 1.0140  & 4.4448 &  315.95 &
5.4503 & 1.0088 & 0.7750 & -2.311 \\ '' (D) &-1.4105 & 4.1146 & 386.99
& 6.5538  &-1.4687 & 0.9997  & -2.376  \\ '' (E)  & 1.2206 &  4.5322 &
418.34  & 5.5932  & 1.2068  & 0.2387  & -2.645  \\ ''  (F)  &-2.0401 &
4.0004& 495.09 & 6.7266 & -2.4071 & 0.9997 & -2.718 \\ '' (G) &-3.9494
& 3.6011&  4520.7 & 7.0388  & -6.9530 &  0.9403 & -2.727 \\  \hline \\
0.169  (A) &-0.7204 &  4.0425 &  480.75& 5.5118  &-0.72463 &  0.9996 &
-1.685 \\ '' (B) &-2.0476 & 3.9250 & 4629.6& 6.3691 &-2.05169 & 1.0000
& -1.875  \\ ''  (C) &-2.3589 &  3.8750 &  11211 & 6.4804  &-2.36304 &
1.0000 & -1.956

\end{tabular}
\medskip

{\small  Ages are counted  from the  end of  mass transfer  (at \teff=
5000K for $M_*$=  0.406 and 0.242 \msun, and  \teff= 10000K for $M_*$=
0.196 and 0.169 \msun)}

\end{minipage}
\end{table*}

\section{Results} \label{sec:results}

In  this section  we describe  the main  results of  our calculations.
Using the evolutionary code  and the treatment for diffusion described
in the  preceding section,  we have calculated  helium WD  models with
stellar masses of 0.406, 0.360,  0.327, 0.292, 0.242, 0.196, 0.169 and
0.161 \msun, which  amply covers the range of  stellar masses expected
for  these  objects.   Realistic  starting  models  were  obtained  by
abstracting  mass from a  1 \msun  \ giant  star up  to the  moment it
begins to  evolve to the  blue part of  the HR diagram. From  then on,
evolution  was followed  down  to very  low  stellar luminosities  and
assuming  a constant value  for the  stellar mass.   As stated  in the
introduction, this  investigation is aimed at  exploring the influence
that diffusion has  on the evolution of helium  WDs. Thus, in addition
to evolutionary  sequences in which  diffusion is considered,  we have
computed sequences for which diffusion is neglected.  This has enabled
us  to  clearly  identify  the  effect induced  by  diffusion  on  the
evolution of  these objects.  We want  to mention that  the results we
have obtained for  the standard treatment of no  diffusion are in good
agreement with those of Driebe et al. (1998).

We begin by  examining Figs 1 to  4 in which we show  the evolution in
the HR diagram of helium WD  models with masses of 0.406, 0.242, 0.196
and  0.169 \msun \  considering (solid  lines) and  neglecting (dotted
lines)  diffusion.   Characteristics  of  the  models  with  diffusion
labeled by letters are given in  Table 2 in which, from left to right,
we list the stellar mass  and photon luminosity (both in solar units),
the  effective temperature, the  age (in  million years),  the surface
gravity ($g$), the nuclear  luminosity (in solar units), the abundance
by mass of surface hydrogen  and the hydrogen envelope mass.  The most
outstanding  feature  illustrated by  these  figures  is that  element
diffusion  plays  a  very  important role  in  inducing  thermonuclear
flashes, which, as we shall see, are critical regarding the subsequent
evolution   of  these   objects  even   during  their   final  cooling
phase. During flash episodes, the star evolves very rapidly and in the
majority of the cases makes an  excursion to the red giant part of the
HR diagram.  Note that if diffusion is neglected, only the 0.242 \msun
model  experiences  thermal   instabilities,  in  agreement  with  the
predictions of Driebe et al.   (1998), who found that hydrogen flashes
occur for two of their sequences:  0.234 and 0.259 \msun \ models.  On
the contrary, we  find that models with masses  greater than $\approx$
0.18 \msun \ experience thermonuclear  flashes when account is made of
diffusion.   This is  true even  for the  most massive  model  we have
analysed here.   This can be understood  on the basis that  there is a
tail  in the  hydrogen distribution  that chemically  diffuses inwards
where  temperature is  high enough  to burn  it, triggering  a thermal
runaway. We shall see that this  will be responsible for the fact that
at late  stages of evolution  ages are very  different in both  set of
calculations.   Note  that  less  massive  models  do  not  experience
hydrogen flashes but they  become markedly less compact when diffusion
is considered.

The effect of  diffusion on the element distributions  within the star
is appreciated in Figs 5 to  7, where the abundances of $^1$H, $^4$He,
$^{12}$C, $^{14}$N and  $^{16}$O are shown as a  function of the outer
mass fraction $q$  ($q= 1- M_r/M_*$) for 0.406,  0.242 and 0.169 \msun
helium  WD models at  various epochs.   In each  case, figure  {\bf a}
shows  the initial  chemical stratification  (as given  by  the pre-WD
evolution) before the  model reaches the cooling branch  for the first
time and  short after  the end  of mass loss  episodes (for  the 0.169
\msun \  model diffusion has  already modified the composition  of the
very  outer layers).   Note that  for the  least massive  model shown,
initial  abundances  in the  outer  layers  are  different from  those
assumed  for  the interstellar  medium.   This  is  because mass  loss
exposed layers  in which hydrogen  burning occurred during  the pre-WD
evolution and essentially all  the initial $^{12}$C was processed into
$^{14}$N.  The effect that  diffusion has on the element distributions
during  the further  evolution is  clearly  noticeable in  all of  the
models analysed  in these figures. Note that  diffusion rapidly causes
hydrogen to float  to the surface, leading to  pure hydrogen envelopes
even in the case of low stellar  masses. At the same time, the tail of
the hydrogen distribution chemically diffuses inward to hotter layers,
making  hydrogen  nuclear  reactions   ignite  there.   This  fact  is
eventually responsible for  the occurrence of additional thermonuclear
flashes.  During  flash episodes,  models develop an  outer convection
zone  thick enough  so  as to  mix  hydrogen and  helium layers,  thus
modifying the composition of the outer layers (see figures {\bf c} and
{\bf e} corresponding to figures 5 and 6, respectively).  After mixing
episodes,  models rapidly  evolve back  to  the red  giant regime  and
finally to the cooling branch (at constant luminosity), at which stage
evolutionary time  scales become longer and diffusion  begins to alter
the chemical  composition again.  This  is in sharp contrast  with the
results  predicted by neglecting  diffusion, in  which case  the outer
layer  chemical composition  is  established by  the  last episode  of
convective mixing and the pre-WD evolution.  Another feature worthy of
comment is that illustrated by figure  {\bf f} of Fig.  6, which shows
that at low luminosities some helium is dredged up to the surface as a
result of convective mixing.  In fact, the hydrogen envelope remaining
after the end of flash episodes is thin enough that convection is able
to mix it with the underlying  helium layers when the star reaches low
luminosities.  Thus, over a  considerable time interval, the star will
be characterized by outer layers  made up of hydrogen and helium. More
specifically, this stage  of mixed envelope lasts for  about 2 Gyr and
it is found to  occur only in the mass interval 0.18  - 0.25 \msun. We
will return to  this point later in this section.   It is important to
remark  that models  without diffusion  also  experience thermonuclear
flashes, but the important point to  be made is that diffusion is able
to carry some hydrogen to inner  and hotter regions and thus to induce
flash episodes even  in models as massive as 0.41  \msun. We shall see
that this fact produces a  different cooling history even at very late
stages of evolution.

The role played by diffusion on  surface gravities can be seen in Figs
8 to  11. As we  mentioned, diffusion causes  the bulk of  hydrogen to
float and  helium and heavier elements  to sink down,  making the star
inflate.  In  addition, the  stellar radius depends  on the  amount of
hydrogen left after flash  episodes, which will be different according
whether diffusion is neglected or  not.  In more massive models and at
high  \teff  values  where   evolution  proceeds  very  fast,  surface
gravities result very different  if diffusion is considered.  At later
stages however,  where evolution is  very slow, surface  gravities are
quite similar.   In fact, because  the inclusion of  diffusion reduces
the  mass  of the  hydrogen  envelope  considerably  (thus leading  to
smaller stellar radius), surface gravities are not appreciably changed
from the situation when diffusion  is neglected.  On the contrary, for
the 0.169  and 0.161 \msun \  models (Fig.  11)  surface gravities are
considerably smaller  when diffusion  is considered.  Here,  models do
not experience thermonuclear flashes even in the presence of diffusion
and evolution  is very  slow during almost  the entire  cooling phase.
The location of the WD companion to the millisecond pulsar PSR J1012 +
5307  according  to Callanan  et  al.   (1998)  determination is  also
indicated in  Fig.  11.   The fact that  models in which  diffusion is
allowed to operate are characterized by lower surface gravities is due
in  part to  that  models  with diffusion  are  characterized by  more
massive hydrogen envelopes (see Fig. 13) because nuclear reactions are
less effective in reducing the  hydrogen envelope of these models.  We
find  for instance that  for models  with $M_*  < $  0.18 \msun  \ the
surface gravity  is reduced  by almost 80  per cent when  diffusion is
considered.  This has the effect  to increase the stellar mass derived
from atmospheric parameters  and mass-radius relations.  Clearly, this
effect should be taken into account when assessing stellar masses from
mass-radius relations  for these objects.  In this  connection, we see
from Fig. 11 that a stellar mass of 0.17 $\pm$ 0.01 \msun \ is derived
for the WD companion to PSR J1012 + 5307 when diffusion is considered,
in contrast  with the  estimation of 0.15  $\pm$ 0.01 \msun  \ derived
from  the assumption  of  no diffusion  (Driebe  et al.  1998 and  our
results without diffusion).

As  mentioned   earlier,  diffusion  is  an   important  mechanism  in
establishing the outer layer composition even in low mass, helium WDs.
In this  context, we show  in Fig. 12  the hydrogen abundance  by mass
($X_H$) at a mass depth of  $10^{-9} M_*$ below the stellar surface as
a function  of age for the  0.196 \msun \ helium  WD sequence.  During
each flash episode, the outer  convection zone digs deep into the star
and reaches helium-rich  layers, thus leading to envelopes  made up by
hydrogen and helium  (for the 0.406 \msun \ model  this takes place at
point  C in  Fig. 1).   At the  last flash  for instance,  the surface
hydrogen  abundance  drops  to  $X_H=$  0.083.   Note  that  the  time
intervals during which the envelope remains helium-enriched are indeed
extremely  short.   In  fact,  the  purity  of  the  outer  layers  is
established by  diffusion as the  star rapidly returns to  the cooling
branch.   However, at  advanced stages  of evolution  and over  a time
interval  of 2  Gyr,  models are  characterized  by helium-rich  outer
layers ($X_H  \approx 0.75$).  This  is because the  hydrogen envelope
remaining  after  the  end  of  flash episodes  is  thin  enough  that
convection is able to mix it  with the underlying helium layers at low
\teff values.  As evolution proceeds, convection reaches the domain of
degeneracy and  begins to retreat outwards (the  maximum depth reached
by the  convection zone  is at $M_{conv}/M_*=  10^{-2.76}$). Diffusion
time scales  at the bottom of  this convection zone  are comparable to
evolutionary  time scales  and  thereby helium  is  depleted from  the
entire  outer  convection  zone  slowly, ultimately  leading  to  pure
hydrogen  envelope again,  as it  is apparent  from Fig.   12.   It is
nevertheless worth mentioning that elements heavier than helium remain
out  of  reach of  the  convection zone  (only  traces  of carbon  are
present, $X_{^{12}C}  \approx 2 \times  10^{-7}$), so the  presence of
metals  in these  stars  should  be explained  basically  in terms  of
accretion episodes.\footnote{Because of the long diffusion time scales
found  at  the base  of  the  convection  zone, metals  accreted  from
interstellar clouds could  be maintained in the outer  layers of cool,
low-mass helium WDs  for a long time, thus  favouring their detection;
see Althaus \& Benvenuto (2000)}.  These results are in sharp contrast
with the situation in which diffusion is neglected.  In this case, the
hydrogen surface abundance remains unchanged ($X_H= 0.504$) throughout
the whole evolution  and it is fixed by the  pre-WD phases during mass
transfer (the  0.196 \msun \  model does not  experience thermonuclear
flashes in  the absence of diffusion).   We mention that  we found the
range  of stellar  masses  for which  the  presence of  helium in  the
envelope is expected at low \teff  to be 0.18 - 0.25 \msun, though the
abundance  of helium  decreases considerably  for more  massive models
(for  instance, the  helium content  reaches $10^{-4}$  for  the 0.242
\msun \ models; see figure {\bf f } of Fig.  6).

\begin{figure}
\epsfxsize=240pt \epsfbox[19 210 578 800]{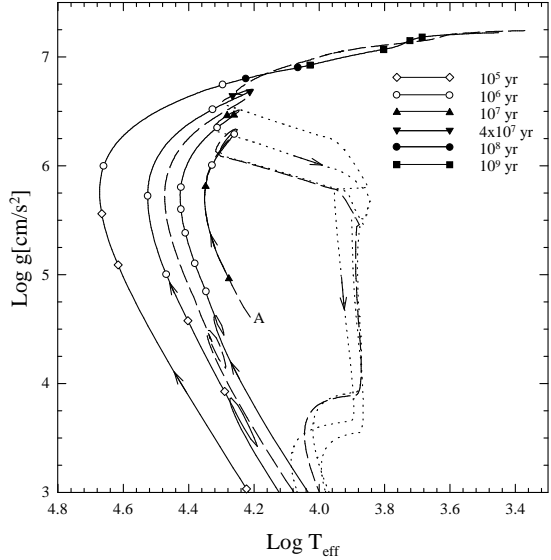}
\vskip -1.2cm
\caption{Same as Fig.   8 but for 0.242 $M_{\odot}$  helium WD models.
Here,  models with diffusion  experience three  thermonuclear flashes,
the effect  of which upon  surface gravities is clearly  noticeable at
high $T_{\rm eff}$ values where evolution proceeds very fast.}
\end{figure}

\begin{figure}
\epsfxsize=240pt \epsfbox[19 210 578 780]{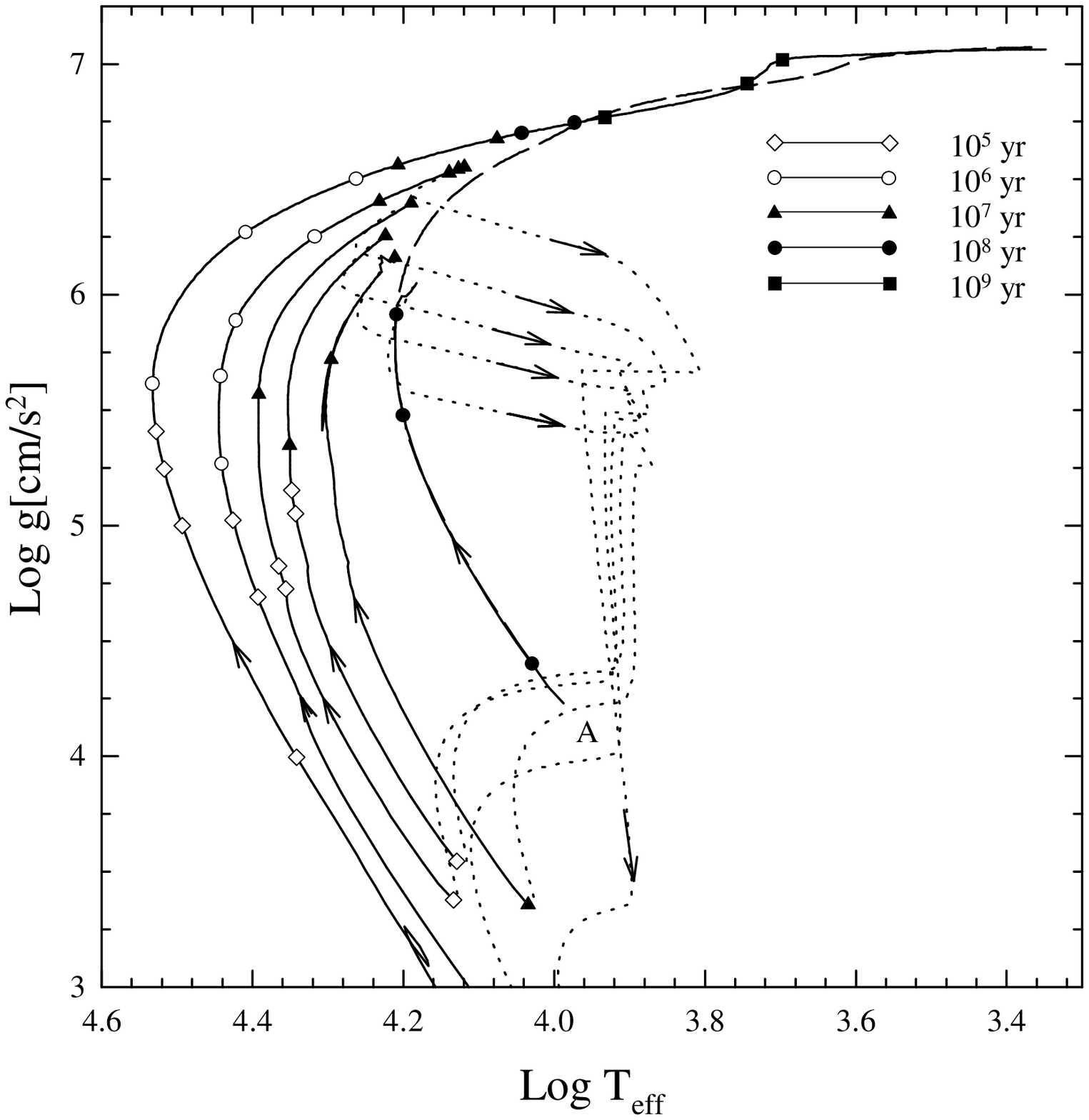}
\vskip -1cm
\caption{Same as Fig.   8 but for 0.196 $M_{\odot}$  helium WD models.
Here, models with diffusion experience five thermonuclear flashes, the
effect of which  upon surface gravities is clearly  noticeable at high
$T_{\rm eff}$ values where evolution proceeds very fast.}
\end{figure}

\begin{figure}
\epsfxsize=240pt \epsfbox[19 210 578 800]{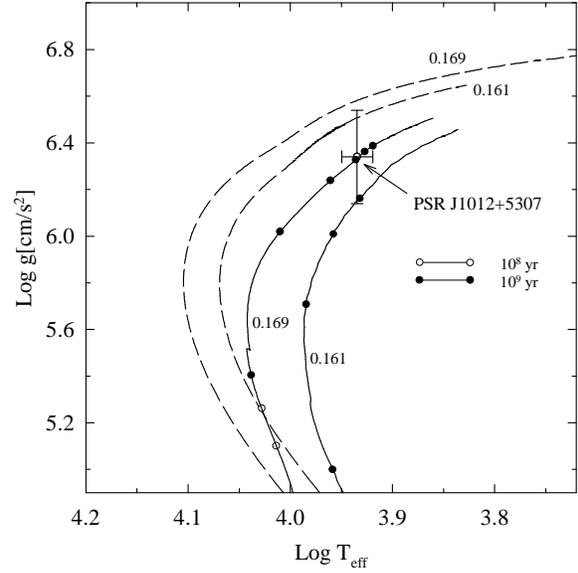}
\vskip -1cm
\caption{Same as Fig. 8 but  for 0.169 and 0.161 $M_{\odot}$ helium WD
models.  Here, models do  not experience thermonuclear flashes even in
the presence  of diffusion.  The location of  the WD companion  to the
millisecond  pulsar PSR  J1012 +  5307  according to  Callanan et  al.
(1998) determination  is also  indicated.  Note that  models in  which
diffusion  is allowed  to operate  are characterized  by substantially
lower surfaces gravities.}
\end{figure}

\begin{figure}
\hskip -.8cm \epsfxsize=260pt \epsfbox[19 210 578 790]{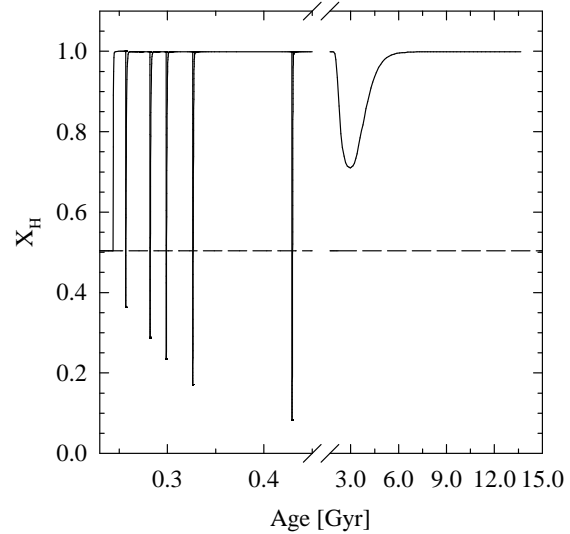}
\vskip -1.7cm
\caption{ Hydrogen abundance by mass (at a mass depth of $10^{-9} M_*$
below the stellar surface) as  a function of age for 0.196 $M_{\odot}$
helium WD models.   Solid line corresponds to the  case with diffusion
and  dashed  line when  diffusion  is  neglected.   During each  flash
episode,  hydrogen  abundance  is  strongly  reduced as  a  result  of
convective  mixing but  the  purity  of the  outer  layers is  rapidly
established by  diffusion. Note that  at advanced stages  of evolution
and over a  time interval of 2 Gyr models  present surface layers made
up of hydrogen  and helium.  In fact, the  hydrogen envelope remaining
after the end of flash episodes is thin enough that convection is able
to mix it with the underlying helium layers. See text for details.}
\end{figure}

\begin{figure}
\epsfxsize=260pt \epsfbox[19 210 578 800]{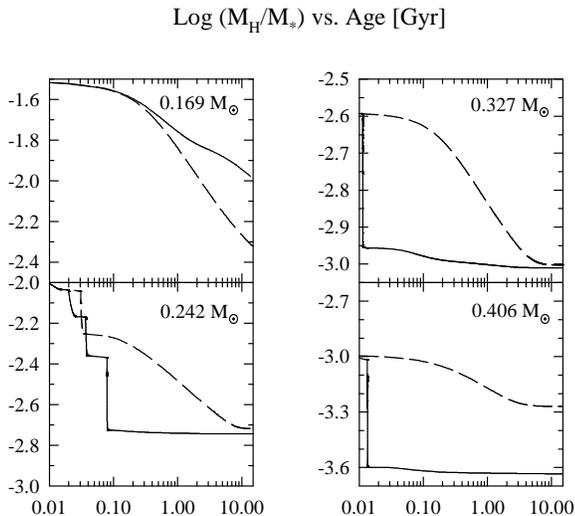}
\vskip -2cm
\caption{ Hydrogen envelope mass against age.  Stellar mass values are
indicated  in each  figure. Solid  line corresponds  to the  case when
diffusion is included and dashed  line to the situation when diffusion
is  neglected. Except  for the  smallest stellar  mass which  does not
experience thermonuclear  flashes, the  amount of hydrogen  left short
after the  end of flash episodes  is markedly lower  when diffusion is
considered. Note  that models  without diffusion burnt  an appreciable
fraction of its hydrogen content on the cooling track.}
\end{figure}

Let us analyse  the amount of hydrogen left atop the  WD after the end
of flash episodes.  This is  an important subject concerning the final
cooling behaviour of  the helium WD.  To this end, we  show in Fig. 13
the mass  of the  hydrogen envelope  as a function  of age  for 0.169,
0.242, 0.327 and 0.406 \msun  \ WD models.  The most important feature
illustrated  by this  figure is  that, except  for the  0.169  \msun \
model, the  amount of hydrogen remaining short  after hydrogen flashes
have  ceased is  markedly lower  when diffusion  is  considered.  More
specifically, in  the case of  models without diffusion,  the hydrogen
mass at entering the final cooling track is about 2 - 4 times as large
as in the case with diffusion.  As mentioned, the 0.242 \msun model is
the  only  one   for  which  we  found  flashes   in  the  absence  of
diffusion.  Note that  in  this situation,  an  appreciable amount  of
hydrogen is  also burnt  during the flash  but the fact  that chemical
diffusion  is able  to  carry  some hydrogen  to  hotter layers  (thus
triggering thermal instabilities) ultimately implies that the sequence
with diffusion must  be left with a much  thinner hydrogen envelope at
entering  the final  cooling  branch.  For  less  massive models,  the
situation is  different because  they do not  experience thermonuclear
flashes  and thus  the  amount of  hydrogen  at the  beginning of  the
cooling  branch  is the  same  irrespective  of  whether diffusion  is
considered or not.   It is noticeable that the  final hydrogen content
for this model is larger  when diffusion is allowed to operate because
diffusion makes the star inflate, leading to lower temperatures at the
bottom  of the envelope  where hydrogen  is burnt.  Here, most  of the
hydrogen content is burnt during the final cooling phase. This is true
even in the presence of  diffusion.  On the contrary, for more massive
models, only when diffusion is not considered an appreciably amount of
hydrogen  is processed  over the  final  evolution. In  view of  these
results, it is expected that the role played by nuclear burning during
the final  WD cooling phase is  different in both sequences,  as it is
indeed borne out by the results  shown in Fig.  14.  Here, we show the
ratio of nuclear  to photon luminosities as a function  of age for the
same stellar masses  as analysed in Fig.  13.   Solid line corresponds
to  the  case  when diffusion  is  included  and  dashed line  to  the
situation  when diffusion  is neglected.   Only the  evolution  on the
final cooling branch is depicted  in the figure.  Except for the 0.169
and 0.161  \msun \  models, for which  nuclear burning is  dominant in
both sequences,  nuclear energy release is negligible  for models with
diffusion.  On the  contrary, in agreement with Driebe  et al.  (1998)
predictions,  for models  without  diffusion hydrogen  burning is  the
dominant energy  source even  at advanced stages  of evolution.  So we
arrive at  a very important  result: {\it diffusion  prevents hydrogen
burning from being  a main source of energy for  most of the evolution
of helium WDs with stellar masses greater than $\approx 0.18$} \msun.

The implications  of the results discussed in  the foregoing paragraph
for the  evolutionary ages are illustrated  by Fig. 15  in which \teff
versus age relation  is shown for all of our  models together with the
observational data  for the WD  companions to the  millisecond pulsars
PSR J1012+5307 and B1855+09 (Callanan  et al. 1998 and van Kerkwijk et
al.   2000, respectively).   Only the  evolution corresponding  to the
final cooling  branch is  depicted in the  figure.  First, we  wish to
emphasize that  our results for  the standard treatment  of neglecting
diffusion  are in  very good  agreement  with the  results derived  by
Driebe et  al.  (1998) on the  basis of the same  assumption.  In this
case, evolution is dictated  by residual hydrogen burning, giving rise
to very  long cooling ages.  In particular,  these evolutionary models
predict an age  for the helium WD in PSR  J1012+5307 in good agreement
with the spin-down  age of the pulsar ($\approx$  7 Gyr). However, for
the WD  companion to  PSR B1855+09 such  models predict ages  above 10
Gyr, in strong discrepancy with the pulsar age of 5 Gyr.

Now, let  us examine  the cooling times  when diffusion is  allowed to
operate.  We saw that in  this case models with stellar masses greater
than $\approx$ 0.18 \msun \  suffer from hydrogen flashes and that the
hydrogen mass  left before they enter  the final cooling  branch is so
small  that   the  nuclear  luminosity  released  by   the  object  is
negligible.   Thereby, the  WD has  to  obtain energy  from its  relic
thermal content,  with the consequent result that  cooling ages become
substantially smaller as compared with  the case in which diffusion is
neglected. This is clearly illustrated  by Fig. 15. In particular, for
the PSR B1855+09 companion, models  with diffusion predict an age of 4
$\pm$ 2 Gyr  in good agreement with the pulsar  age.  For masses lower
than  0.18  \msun \  nuclear  burning  is  dominant and  cooling  ages
resemble   those  derived   from  models   without   diffusion.  Thus,
evolutionary  models  taking into  account  element diffusion  predict
cooling  ages  consistent  with   the  ages  of  the  above  mentioned
pulsars. This  naturally solves the  age discrepancy for  PSR B1855+09
system, making it unnecessary to  invoke ad-hoc mass loss episodes, as
proposed recently by Sch\"onberner, Driebe and Bl\"ocker (2000).

\begin{figure}[t]
\epsfxsize=260pt \epsfbox[19 210 578 785]{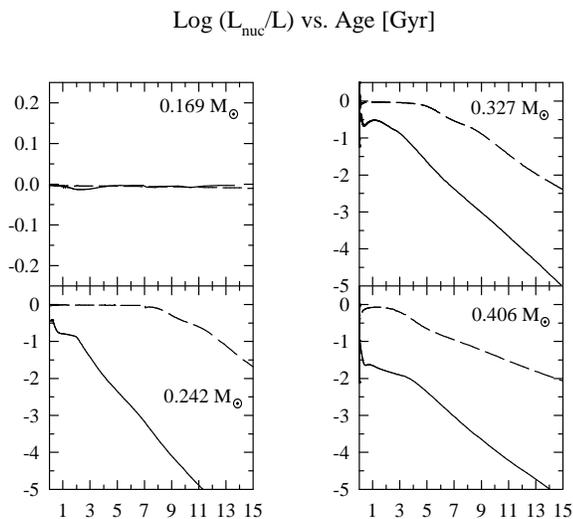}
\vskip -2cm
\caption{Ratio of nuclear to photon luminosities as a function of age.
Stellar  mass  values  are   indicated  in  each  figure.  Solid  line
corresponds to the case when  diffusion is included and dashed line to
the  situation when diffusion  is neglected.  Except for  the smallest
stellar mass which does  not experience thermonuclear flashes, nuclear
energy release is negligible for  models with diffusion on the cooling
branch.  However,  for models  without diffusion, hydrogen  burning is
the dominant energy source even  at advanced stages of evolution, with
the consequent result that evolution is considerably slowed down. }
\end{figure}

\begin{figure}[t]
\epsfxsize=260pt \epsfbox[19 210 578 800]{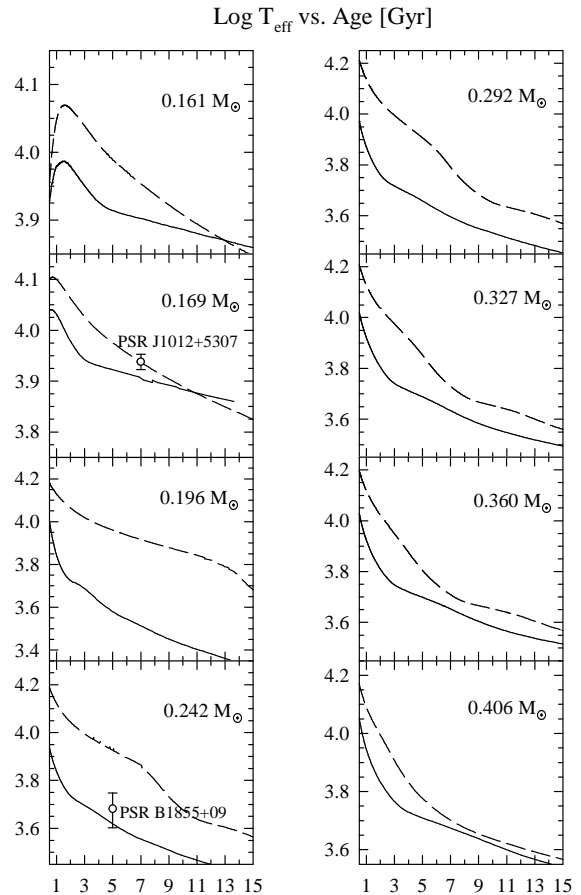}
\vskip 2cm
\caption{Effective temperature as a function  of age for the whole set
of our  models. Solid line corresponds  to the case  when diffusion is
included and dashed line to the situation when diffusion is neglected.
The observational data for the WD companion to the millisecond pulsars
PSR B1855+09 and PSR J1012+5307 are included.  At low $T_{eff}$ values
and for stellar masses greater than $\approx$ 0.18 $M_{\odot}$, models
with  diffusion  evolve  significantly  faster as  compared  with  the
standard  treatment  of  no   diffusion.   Note  that  models  without
diffusion predict ages in good agreement  with the spin - down age for
PSR  J1012+5307  but not  for  PSR  B1855+09.  On the  contrary,  when
diffusion is considered  cooling ages are consistent with  the ages of
both pulsars.}
\end{figure}

\section{Discussion and conclusions} \label{sec:conclusion}

Motivated by  recent observational data  of low-mass white  dwarf (WD)
companions to  millisecond pulsar (van  Kerkwijk et al.   2000), which
seem  to  cast doubts  on  the  thickness  of the  hydrogen  envelopes
predicted by  standard evolutionary calculations (Driebe  et al. 1998;
Sarna et  al. 1999), we  have undertaken the current  investigation in
order to assess the role played by element diffusion in the occurrence
of hydrogen flashes in helium  WDs and more importantly to investigate
whether or not the hydrogen  envelope mass can be considerably reduced
by enhanced  hydrogen burning during  flash episodes. To this  end, we
have  explored the  evolution of  0.406, 0.360,  0.327,  0.292, 0.242,
0.196, 0.169 and  0.161 \msun \ helium WD  models in a self-consistent
way with the evolution  of element abundances resulting from diffusion
processes. The diffusion calculations  are based on the multicomponent
treatment  of the gas  developed by  Burgers (1969)  and gravitational
settling, and  thermal and chemical diffusion have  been considered in
our  calculations.   Reliable initial  models  have  been obtained  by
abstracting mass to a 1 \msun  \ model at appropriate stages along its
red  giant  branch  evolution.   In  the  interests  of  a  consistent
comparison,  the  same  stellar  masses were  evolved  but  neglecting
diffusion. In this regard, our  results are similar to those of Driebe
et al. (1998) on the basis  of the same assumption. We want to mention
that in the  calculations presented in this paper  we have not invoked
additional mass transfer when models return to the red giant region of
the HR diagram as a result  of a thermonuclear flash. In fact, the aim
of the  work has been to  assess the possibility  that diffusion alone
can eventually lead to hydrogen  envelope thin enough so as to prevent
nuclear burning from  being an important source of  energy at advanced
cooling stages.   In this  sense, our treatment  differs from  that of
Iben \& Tutukov (1986) for the case  of a 0.3 \msun \ model.  In their
model,  high mass  loss causes  a markedly  reduction of  the hydrogen
envelope mass  and this  is responsible for  the fact that  the object
cools down rapidly during the final cooling phase.

From  the results  presented in  this work  it is  clear  that element
diffusion considerably  affects the structure and  evolution of helium
WDs.   In fact,  diffusion produces  a different  cooling  history for
helium WDs  depending on the mass  of the models.  We  have found that
models with stellar masses ranging  from $\approx$ 0.18 \msun \ to the
most massive  model we analyse  (0.406 \msun) experience at  least one
thermonuclear flash  when account  is made of  diffusion.  This  is in
sharp contrast  with the predictions  of the standard treatment  of no
diffusion for  which only  the sequence with  0.242 \msun  \ undergoes
flash  episodes   (note  that  Driebe  et  al.    1998  found  thermal
instabilities only  for their 0.259  and 0.234 models).  In connection
with the  further evolution,  these diffusion-induced flashes  lead to
much  thinner hydrogen  envelopes, preventing  stable  nuclear burning
from being a sizeable energy source. Because the star has a much lower
amount of available energy,  it implies much shorter evolutionary time
scales as compared with the  situation when diffusion is neglected, in
which  case evolution  is  delayed to  very  long ages  by the  active
hydrogen burning zone that characterizes such models.

These  results have  important implications  when attempt  is  made in
comparing   theoretical   predictions  on   the   WD  evolution   with
expectations from millisecond  pulsars. This is particularly important
in view of recent observational data on the helium WD companion to the
pulsar PSR B1855+09.  In fact,  van Kerkwijk et al.  (2000) determined
the \teff of  the WD companion to be 4800 $\pm$  800K.  Since the mass
of  the WD  is accurately  known thanks  to the  Shapiro delay  of the
pulsar signal  (0.258$^{+0.028}_{-0.016}$\msun), this low  \teff value
corresponds to a  WD cooling age of 10 Gyr according  to the Driebe et
al. evolutionary models  or our models without diffusion.   This is in
strong discrepancy with  the spin-down age of the  pulsar (5 Gyr).  On
the contrary, we  found that when diffusion is  properly accounted for
in stellar  models of helium WD,  evolution is accelerated  to such an
extent  that this age  discrepancy vanishes  completely.  In  fact, we
found that  our models  predict an  age of 4  $\pm$ 2  Gyr for  the WD
companion to  PSR B1855+09. This naturally solves  the age discrepancy
for PSR  B1855+09 system, making  unnecessary invoke ad-hoc  mass loss
episodes, as  proposed recently by  Sch\"onberner et al. (2000),  or a
non standard braking  index. We also found that  for masses lower than
0.18  \msun, nuclear  burning  is  dominant even  in  the presence  of
diffusion and cooling ages  resemble those derived from models without
diffusion.

In light of  the new results presented in this  work, it is worthwhile
to re-examine other millisecond  pulsar systems with WD companions for
which  cooling  and  spin-down  ages  appear to  be  discrepant.   PSR
J0034-0534 system is particularly  noteworthy in this regard. In fact,
the WD in this  system is very cool (\teff $< $  3500K) and the pulsar
age  is 6.8  $\pm$ 2.4  Gyr (Hansen  \& Phinney  1998b).  The  WD mass
according to the relation between the orbital period of the system and
the WD  mass (Tauris  \& Savonije 1999)  results $\approx$  0.21 \msun
(see  Sch\"onberner et  al.   2000).  Helium  WD  models that  neglect
diffusion predict ages well above 10 Gry for this WD companion. On the
contrary, evolutionary  calculations including element  diffusion give
for this  helium WD an  age of 6  and 8 Gyr  at \teff= 3500  and 3000,
respectively, which is in excellent  agreement with the pulsar age. It
is clear that consistency between  pulsar and WD ages in PSR J0034-053
system can  be obtained  with helium  WD models and  not only  with WD
models having a carbon-oxygen core  with stellar mass of 0.5 \msun, as
argued by  Sch\"onberner et al.   (2000) on the basis  of evolutionary
models in which  diffusion is neglected. Note that the  mass of the WD
derived by Sch\"onberner et al.  is clearly at odds with that inferred
from the relation of Tauris \& Savonije (1999).

Other millisecond pulsar system with  a cool WD companion of relevance
in the context  of the new results presented here and  which is also a
discrepant  case is  PSR J1713+0747.   For  the WD  companion to  this
pulsar,  Hansen \& Phinney  (1998b) give  an effective  temperature of
\teff $<$  3800K and a spin-down age  for the pulsar of  9.2 $\pm$ 0.4
Gyr.   The WD mass  according to  the relation  of Tauris  \& Savonije
(1999) is $\approx$ 0.33 \msun.  Models that neglect diffusion predict
an age  for the  WD greater than  13 Gyr  at that \teff  value, whilst
models with  diffusion give an age  of 9.1 Gyr in  very good agreement
with the pulsar age.

In closing, it is clear  that element diffusion plays a very important
role in  the cooling history of  helium WDs.  From the  results of the
present  study,  we  conclude   that  age  discrepancies  between  the
predictions   of  standard   evolutionary   calculations  and   recent
observational data  of millisecond  pulsar systems with  WD companions
appear  to be  the result  of ignoring  element diffusion  in existing
evolutionary  calculations.   Indeed,  such  discrepancies  completely
vanish when account is made of diffusion.

Finally, we  wish to  mention that because  of the employment  of gray
atmospheres in our calculation, cooling  ages at very low \teff values
should be  taken with  some caution. As  discussed recently  by Hansen
(1999) and Salaris et al.  (2000) in the context of carbon-oxygen WDs,
an adequate treatment of the atmosphere is required at advanced stages
of evolution  where the gray assumption becomes  a poor approximation.
We are planning to include  in our helium WD evolutionary calculations
a non gray  atmosphere as an outer boundary  condition appropriate for
very old helium WDs.

Complete tables containing the results of the present calculations are
available  at   http://www.fcaglp.unlp.edu.ar/$\sim$althaus/  or  upon
request to the authors at their e-mail addresses

\bsp
\label{lastpage}
\end{document}